\documentclass[manuscript]{aastex}

\shorttitle{Statistical Classification to the Globular Clusters}
\shortauthors{Tanuka Chattopadhyay \& Pradip Karmakar}

\begin{document}

\title{Statistical analysis of dwarf galaxies and their globular clusters in the Local Volume}

\author{Tanuka Chattopadhyay,\altaffilmark{1}}
\affil{$^1$Department of Applied Mathematics, Calcutta
            University,
             92 A.P.C. Road, Calcutta 700009, India} \email{tanuka@iucaa.ernet.in}

\author{Margarita Sharina\altaffilmark{2,3,4}}
\affil{$^2$Special Astrophysical Observatory, Russian Academy of
Sciences, N. Arkhyz, KCh R, 369167, Russia} \affil{$^3$
Laboratoire d'Astrophysique de Toulouse-Tarbes, Universit\'e de
Toulouse, CNRS, 14 avenue E. Belin, F-31400, France}
\affil{$^4$Isaac Newton Institute, Chile, SAO Branch}
 \and

\author{Pradip Karmakar\altaffilmark{1}}
\affil{$^1$Department of Applied Mathematics, Calcutta
            University,
             92 A.P.C. Road, Calcutta 700009, India}

\begin{abstract}

Morphological classification of dwarf galaxies into early and late
type, though can account for some of their origin and
characteristics but does not help to study their formation
mechanism. So an objective classification using Principal
Component analysis together with K means Cluster Analysis of these
dwarf galaxies and their globular clusters is carried out to
overcome this problem. It is found that the classification of
dwarf galaxies in the Local Volume is irrespective of their
morphological indices. The more massive ($M_{V0} < -13.7$)
galaxies evolve through self-enrichment and harbor dynamically
less evolved younger globular clusters (GCs) whereas fainter
galaxies ($M_{V0} > -13.7$) are influenced by their environment in
the star formation process. 
\end{abstract}

\section{Introduction}

The galaxies with low luminosities, low metallicities having
smaller sizes are termed as dwarf galaxies. Study of dwarf
galaxies is important as massive galaxies are supposed to be
formed by hierarchical merging of dwarf galaxies during the
evolution of early Universe (White \& Rees 1978; Geisler et al.
2007; Haines et al. 2006). Such objects lose gas easily due to
their shallow potential wells. Low surface brightness (LSB)
 dwarf galaxies are classified primarily into
three groups : early type ( dwarf spheroidal, dSph, and dwarf
elliptical, dE), late type (dwarf irregular, dIrr), and transition
type galaxies (Kormendy 1985; Karachentseva et al. 1985; Grebel
1999). There is no sharp boarder line between these morphological
types (see e.g. Sharina et al. 2008 and references therein,
hereafter: S08).
 Also population gradients exist in
early-type dwarf galaxies, and dSphs and dIrrs have exponential
surface brightness profiles. All these facts indicate that
classification, based on morphology and stellar content is not
sufficient to study the formation and evolutionary status of these
objects. A more sophisticated classification is essential for
studying true formation mechanisms of this class of objects.
However the objects under consideration share one common property.
They all harbor globular clusters those are older than several Gyr
which indicates that early globular cluster formation took place
irrespective of the morphological type. So these globular clusters
can serve as unique tool to investigate the chemical evolution of
the host galaxies. Hence a proper classification of globular
clusters in LSB galaxies is necessary for finding information
regarding star formation histories in these dwarf galaxies which
is important input for studying galaxy formation mechanism. Since
Hubble (1922, 1926) tunning fork diagram, very little attempts
have been taken on  development of an objective classification for
normal and dwarf galaxies using statistical methods and principles
(Whitmore 1984; Vaduvescu \& McCall 2005; Fraix Burnet et al.
2006; Chattopadhyay \& Chattopadhyay 2006; van den Bergh 2007; van
den Bergh 2008; Woo et al. 2008) though few works have been
carried out on dE classification and formation scenarios
(Marin-Franch \& Aparicio 2002; Lisker et al. 2007; Penny \&
Conselice 2008) in Virgo, Perseus and Coma clusters of galaxies
which is again the consideration of a particular morphological
type and therefore not exhaustive. In order to identify the
parameters those are mostly responsible for the variation among
the dwarf galaxies and their globular clusters and to classify
them into homogeneous groups for searching the possible formation
mechanism we have to use some statistical techniques like
Principal Component Analysis
(PCA) and Cluster Analysis (CA).\\

In the present paper, in order to study the underlying features of
the dwarf galaxy population we have used statistical methods like
PCA, CA and Discriminant analysis. By treating the samples under
consideration as representatives of the corresponding underlying
population of dwarf galaxies , these methods help us to make
inference regarding the above mentioned population ( and not only
for the samples under consideration). As a result, on the basis of
the present study we can make some general conclusions which are
not feasible on the basis of visual studies.

In Section 2 the data sets are discussed. The methods, used, are
described in Section 3 while results, discussions and conclusions
are summarized in Sections 4, 5 and 6 respectively.

\section{Data Set}

Our analysis is based on two data sets of dwarf galaxies and their
globular clusters (GCs) in the Local Volume
(LV).\\
             {\bf Data set 1}\\
              This consists of 60 dwarf
             galaxies taken from a data set of 104 dwarf galaxies
             (Sharina et al. 2008) (Table 1).
             The parameters considered
             from Sharina et al. (2008 hereafter S08) are distance modulus ($\mu_0$, in mag),
             morphological index (T), mean metallicity of the red giant branches ([Fe/H],
             in dex), effective color corrected for extinction($(V-I)_{e0}$, in
             mag), logarithm of projected major axis from CNG($log(Diam)$, in Kpc),
             logarithm of limiting diameter ($log(Dlim)$, in Kpc), limiting V and I
             absolute magnitudes within the diameter Dlim corrected for extinction ($M_{V0}, M_{I0}$),
             extinction corrected mean SB within 25 magnitude isophote in V and I magnitudes
              ($SBV_{25,0}, SBI_{25,0}$ in $mag  arcsec^{-2}$),effective surface
             brightness in V band corrected for extinction ($SBV_{e0}$, in $mag arcsec^{-2}$),
             logarithm of effective radius ($log(R_e)$ in Kpc), logarithm of
             model exponential scale length ($logh$, in Kpc), best exponential fitting
              central surface brightness in V and I bands corrected for extinction (SBVC0,
              SBIC0 in $mag arcsec^{-2}$) respectively.
              The parameters used from Karachentsev et al.(2004, hereafter: CNG)
              are HI rotational velocity ($V_m$ in $Kms^{-1}$), HI
              mass to luminosity ratio ($M_{HI}/L$ in solar units),
              and tidal index ($\Theta$). The scaling parameters used from Georgiev
              et al. (2010) are globular cluster specific
              frequency ($S_N$), specific luminosity ($S_L$),
              specific mass ($S_M$), specific number ($\hat{T}$),
              logarithm of specific GC formation efficiency as a
              function of galaxy luminosity and mass ($\eta_L,
              \eta_M$), total stellar mass
              ($M_{*,V}$ in $10^7 M_{\odot}$) and HI mass of the host
              galaxy ($M_{HI}$ in $10^7 M_{\odot}$) respectively. \\
              During selection of parameters for PCA and CA the following things were
             taken into consideration.\\
             (i) The parameters must be intrinsic in nature.\\
             (ii) For almost physically similar parameters any one
             is chosen at random because inclusion of similar
             parameters are considered as redundant in CA.\\
             (iii) All the parameters should be without missing
             values as CA does not allow parameters having missing
             values exceeding 5\% for which mean substitution
             might be allowed (Little \& Rubin 2002).\\
             With respect to the above aspects (viz. (i)) we excluded $\mu_0$
             and T which were not intrinsic properties of dwarfs.
             We included $ SBV_{25,0}, SBI_{25,0}, M_{V0}, M_{I0}, logR_e, logh$. So
             $SBVL, SBIL, MV_{25}, MI_{25}, R_{V,25},
             R_{I,25}$ were not included with respect to (ii). We
             have not considered remaining parameters except $\Theta$ with respect
             to (iii) as they have missing values exceeding 5\%
             but once the dwarf galaxies are classified we used
             them to study their properties in more detail.
             So among all these parameters only 13 parameters from Sharina et
             al. (2008)excluding $\mu_0$ and T, together with ($\Theta$)
             from Karachentsev et al.(2004) are directly used for
              PCA and CA as the
              sample is  without any missing values with
              respect to these 14 parameters. This is a very standard procedure
              followed during PCA and CA for a sample of astronomical objects.
              In order to have a sample where the values of all the parameters corresponding
              to each dwarf galaxy are available, we had to drop observations
              corresponding to remaining 44 dwarf galaxies and as a result we get a sample of size 60
              from the original one.  The sample is not complete as there are many more galaxies in the LV
              which are yet to be observed. In this sense no catalogue is complete. The question remains
              whether the sample is a good representative of the original one or not.
              Regarding this point all the two point correlations discussed in the previous paper
              (S08) are still in place after the selection of the present sample.
              It is important, that it contains all transitional type galaxies
             (dSph/dIrr, T = -1) from the original sample.
              So all morphological types are well represented in this sense. A list
              of dwarf galaxies considered in Data set 1
              is given in Table 1.

              {\bf Data set 2}\\
             This consists of 100 GCs in the Local Volume dwarf galaxies (Sharina et al. 2005).
             Three candidates Sc 22-2-879, Sc 22-100 and Sc
             22-4-106 were removed as they are identified later as
             galaxies and not GCs (Da Costa et al. 2009). Also the parameters of the GCs
             in UGC4115, KK65 and UGC3755 are recalculated using
             the current distances 7.727, 8.017 and 7.413 Mpc
             repectively (Tully et al. 2006).
             The parameter set consists of logarithm of half light radius
             ($log(r_h)$ in parsec), apparent axial ratio (e),
             integrated absolute magnitude (V0, in mag)
             corrected for extinction, integrated absolute $(V-I)_0$
             color (corrected for Galactic extinction, in mag),
             projected distance from the host galaxy ($d_{proj}$,
             in Kpc), central surface brightness in V and I bands
             ($\mu_{V0}$, $\mu_{I0}$ in $mag arcsec^{-2}$), logarithm of  King core
             radius and tidal radius ( $log(r_c)$, $log(r_t)$ in
             parsec) respectively.\\

\section{Method}

Principal Component Analysis (PCA) is a very common technique used
in data reduction and interpretation in multivariate analysis. We
are interested in discovering which parameters in a data set form
coherent subgroups that are relatively independent of each other.
The specific aim of the analysis is to reduce a large number of
parameters to a smaller number while retaining maximum spread
among experimental units. The analysis therefore helps us to
determine the optimum set of parameters causing the overall
variations in the nature of objects under consideration. PCA has
been discussed and used by various authors (Babu et al. 2009;
Chattopadhyay \& Chattopadhyay 2006, 2007; Whitmore 1984; Murtagh
\& Heck 1987).\\

Cluster analysis  (CA) is the art of finding groups in data. Over
the last forty years different algorithms and computer programs
have been developed for CA. The choice of a clustering algorithm
depends both on the type of data available
and on the particular purpose. \\

 In the present study we have used K- Means partitioning algorithm
 (MacQueen 1967)for clustering. This method  constructs K clusters
 i.e. it classifies the data into K groups which together satisfy
 the requirement of a partition such that each group must
 contain at least one object and each object must belong to exactly one group.
 So there are at most as many groups as there are objects ($K <=n$).
  Two different clusters cannot have any object in common and the K groups
 together add up to the full data set. Partitioning methods are
 applied if one wants to classify the objects into K clusters
 where K is fixed (which should be selected optimally). The aim
 is usually to uncover a structure that is already present in the
 data. The K- Means is probably the most widely applied partitioning
 clustering technique.\\

  Here to perform K-means clustering we have used MINITAB package.
  Under this package cluster centers have been chosen on the basis
  of group average method which makes the process almost robust.
  This method has been developed by Milligan (1980).\\

  By using this algorithm we first determined the structures of sub
populations (clusters) for varying numbers of clusters taking
K=2,3,4 etc. For each such cluster formation we computed the
values of a distance measure $d_{K} = (1/p) min_{x} E[( x_{K} -
c_{K} )^{'}(x_{K} - c_{K})]$ which is defined as the distance of
the $x_{K}$ vector (values of the parameters) from the center
$c_{K}$ (which is estimated as the mean value), p is the order of
the $x_{K}$ vector. Then the algorithm for determining the optimum
number of clusters is as follows (Sugar \& James 2003). Let us
denote by $d_{K}^{\bf'}$ the estimate of $d_{K}$ at the $K^{th}$
point. Then $d_{K}^{\bf'}$ is the minimum achievable distortion
associated with fitting K centers to the data. A natural way of
choosing the number of clusters is to plot $d_{K}^{\bf'}$ versus K
and look for the resulting distortion curve (Figs.1 \& 2, bottom
one of each figure). This curve is always monotonic decreasing.
Initially one would expect much smaller drops for K greater than
the true number of clusters because past this point adding more
centers simply partitions within groups rather than between
groups. According to Sugar \& James (2003), for a large number of
items the distortion curve when transformed to an appropriate
negative power (p/2), will exhibit a sharp "jump" (if we plot K
versus transformed $d_{K}^{\bf'}$). Then we calculated the jumps
in the transformed distortion as
 $J_{K} = (d_{K}^{\bf'-{p/2}}  - d_{K-1}^{\bf'-{p/2}}$).

The optimum number of clusters is the value of  K associated with
the largest jump. The largest jump can be determined by plotting
$J_{K}$ against K and the highest peak will correspond to the
largest jump (Figs.1 \& 2, top one of each figure).\\

  It is well known that both the methods PCA and CA are parameter
dependent and the parameters considered should be responsible for
the variation of the objects under consideration. In the present
situation all the parameters of that type are taken into
consideration. As we have to depend on the available data only, it
was not possible for us to consider many unobserved parameters
whose inclusion might have improved the classification e.g.
inclusion of central velocity dispersion etc. and many more. But
the question is, given the parameters and sample whether the
classification is robust or not. In this respect a discriminant
analysis is performed ( Johnson \& Wichern 1998) to verify the
acceptability of the classification by computing misclassification
probabilities for the different dwarfs and GCs. If the original
classification is robust then every dwarf or GC should be
classified again as a member of the same class that it was before.
Tables 2, 3 show the result of a discriminant analysis.The
fractions of correct classifications are 0.983 and 0.97
respectively which imply that the classifications are almost
robust. As in the present situation we have only one sample, it is
difficult to say whether the same results will be obtained for
other samples also. It can only be inferred that if the present
sample is a good representative of the underlying population of
the dwarf galaxies, then the results obtained in this paper are
generally true.

\section{Dwarf galaxies of the Local Volume }

For PCA, at first we have computed a correlation matrix with all
the 14 parameters for Data set 1 and have taken any one of the two
physically similar (e.g. absolute magnitudes in V and I bands)
highly correlated (correlation $>$ 0.7) parameters. Following this
method 8 parameters are selected for PCA. They are $ \Theta,
[Fe/H], M_{V0}, SBV_{25,0}, (V-I)_{e0}, SBV_{e0}, log(R_e), logh $
respectively. For these 8 parameters, PCA analysis gives four
Principal Components with eigen values greater than or equal to 1
and at the same time almost 87.7 \% overall variation. So we have
taken these four Principal Components and have computed the
correlations of the parameters appearing in each Principal
Component with the corresponding Principal Component. We have
considered those parameters as significant one for which the
correlation is greater than 0.65 as a thumb rule. Thus following
this procedure the significant parameters as outcome are $M_{V0},
SBV_{e0}$ and $log(R_e)$ (from the first Principal Component),
$\Theta, (V-I)_{e0}$ (from the second one) and $[Fe/H]$(from the
third one). Fourth component
contribute no parameters with such a high correlation.\\

For cluster analysis we have taken the above six significant
parameters and used the method assuming K = 1,2,3 etc. The optimum
number of coherent groups by the above method is obtained at K=2
(viz. G1 and G2). The 'distortion' and the 'jump' curves are shown
in Fig. 1. The mean values with standard errors for some
parameters and significant correlations with their p values are
shown in Table 4 for the groups G1 and G2 respectively.\\

Under the multivariate situation the role of all the parameters
are important for classification as they are correlated to each
other but sometimes one or two parameters may play a significant
role over the others when there are large variations among the
values of those parameters. In the present situation the magnitude
($M_{V0}$), tidal index ($\Theta$) and effective surface
brightness ($SBV_{e0}$) play such role (Table 3). Although in
terms of magnitude it is possible to find a single cut at $M_{V0}
= -13.7$ irrespective of the other two (Fig. 3), if we consider
Fig. 4 it is clear that no such single cut is available in terms
of $\Theta$. As such the classification is based mainly on the
three major parameters $M_{V0}, \Theta$ and $SBV_{e0}$ and not
only the magnitude. In such multivariate set up we discuss the
marginal situations (i.e. the effect of some single parameter) in
order to display the results graphically so that one can visualize
the underlying scenario. e.g. Fig.3 is a two dimensional
projection of the six dimensional
original situation.\\

 Further on the basis of PCA and CA we have divided the objects
 into certain groups with respect to certain parameters. The
 parameter ranges for different groups are dependent on one
 another and it depends on various factors like range of the
 parameters, size of the sample etc. Hence the feature that on the
 basis of the magnitude less than or greater than -13.7 we can get
 two different groups is not necessarily always true. But the
 feature which is likely to be retained for different samples is
 that in most of the situations there will be two significant
 classes whose distributional natures are likely to be the same as
 that of the present situation. In case where the ranges of
 the sample parameters will be close to the present situation then
 one may expect the similar cut in the value of the magnitude.

\subsection{Globular clusters in the dwarf galaxies of Local Volume}
For PCA, we have taken the parameters $logr_c, logr_h, e, V0,
(V-I)_0, d_{proj}, \mu_{I0}, \mu_{V0}, logr_t$. The computed
correlation matrix does not show high correlation for physically
similar parameters. So all these 9 parameters have been considered
for PCA. The number of principal components with eigen values
close to 1 is 4 for total variation of 83.3 \%. For these 4
principal components very high correlations occur only for two
parameters so we have considered correlations having values
greater than 0.6 as a thumb rule. Following this the significant
parameters are $ logr_c, logr_h, e, V0, (V-I)_0, \mu_{I0},
\mu_{V0}, logr_t$. Next a CA is carried out with these eight
parameters (standardized) and the optimum number of classes is
found to be at K = 4. The 'distortion' and 'jump' curves are shown
in Fig. 2. The mean values for some parameters are listed in Table
5.

\section{Discussions}
\subsection{Dwarf galaxies}
Two groups G1 and G2 of dwarf galaxies in the LV have been found
as a result of CA, which are irrespective of their morphological
classification (viz. T). In G1 3\% are dIrr/dSphs and 97\% are
dIrrs whereas in G2 52\% are dIrrs, 43 \%  are dSphs, and 5\% are
dIrr/dSphs (1 galaxy).  The groups have many distinct properties
as seen from Table 4. G1 contains brighter galaxies of larger size
with larger amount of HI mass having high degree of rotation
whereas G2 consists of fainter galaxies of smaller size and are
almost devoid of HI mass having insignificant amount of rotation.
A luminosity - metallicity (viz. $ [Fe/H] vs M_{V0}$) diagram
(Fig.3) shows a significant correlation (viz. Table 2, $ r\sim
-0.553, p = 0.001$; 2 galaxies on top were removed as outliers)
together with the best fitted line for the galaxies in G1. The
slope of this relation is identical to the one found for dSphs and
dIrrs in the Local Group and beyond (Dekel and Silk 1985; Skillman
et al.; 1989; Smith 1985; S08). Just the zero point is shifted by
$\sim$4 mag. Note, that if we consider the G2 in total, such
correlation is absent (viz. Table 2, $ r\sim -0.290, p = 0.160$).
This may indicate that formation of dwarf galaxies is governed by
self enrichment whereas some processes lead to the fading during
formation  and evolution of stars in them, and interaction of
interstellar gas of dwarf galaxies with with intergalactic medium
in groups (see e.g. Grebel et al. 2003 and references therein).
Gravitational potentials are not strong, and gas may be blown out
by just few supernovae. Galactic winds lead to a significant loss
of metals from dwarf galaxies. Starvation (Shaya \& Tully 1984),
tidal, or ram pressure gas stripping affect galaxies in dense
galaxy group, or cluster environments. The complex behavior of the
liminosity - metallicity in the G1 and G2 also might be accounted
by multiple bursts of star formation of short duration in dwarf
galaxies of small sizes (Carraro et al. 2001; Hirashita et al.
2000). The presence of HI rotation in G1, and almost complete
absence of gas
in G2 supports the above picture.\\

 Fig.4 shows the tidal index
vs. logarithm of the scale length for the sample galaxies. The
so-called ``tidal index'' was introduced by Karachentsev \&
Makarov (1998).
 It is the maximum
logarithm of the local mass densities produced by neighbours of a
galaxy. It is seen that for galaxies with tidal index larger than
zero scale lengths grow with the growing of the tidal index. This
means that neighbours influence the thickening of galactic disks
irrespective of morphological types. G2 is more affected by tidal
interaction, than G1.\\

Fig.5 shows absolute magnitude vs. logarithm of the scale length
for the sample galaxies. Dashed line indicates $h \sim L^{0.5} $
relation for spiral galaxies. It is seen that the slope of this
relation does not change at $M_v \sim -12$ mag as it was suggested
by S08. We see two sequences of galaxies, well divided on the two
groups found in our paper. The shift between the two sequences is
about 2 magnitudes along the X direction, which is as twice as
less in comparison to the luminosity -- metalliicty relation.\\

 One may suggest, the shift in magnitudes between G1 and G2 at
the same metallicity (Fig. 3) and at the same scale length (Fig.
5) is driven by interplay of different factors. The thickening of
disks is produced by interaction with neighbors (tidal, ram
pressure stripping) and by disruption of star clusters (Kroupa
2002). The luminosity -- metallicity relation is the result of the
aforementioned reasons plus effects of stellar evolution. Since we
see the parallel shift according to the absolute magnitude in
Fig.3 and 5, one may conclude, that G2, which contains all dSphs,
evolved from G1 due to the many reasons, such as: fading due to
cessation of star formation gas outflows produced by supernovae,
ram pressure and tidal stripping.

\subsection{Globular cluster candidates}
 Georgiev et al.(2010) have given a conjecture of the formation history of
globular clusters in dwarf galaxies on the basis of stellar and
galaxy mass. They investigated the formation of GCs in terms of
some observed scaling parameters which were theoretically
predicted as a function of galaxy mass on the basis of a model by
Dekel \& Birnboim (2006). These scaling parameters are specific
frequency ($S_N = N_{GC}\times 10^{0.4(M_V+15)}$ where $N_{GC}$ is
the number of GCs and $M_V$ is the absolute visual magnitude of
the host galaxy), specific mass ($S_M = 100\times
M_{GCS}/(M_*+M_{HI})$, where $M_{GCS}$ is the total mass of GCs,
$M_*$ is the total stellar mass and $M_{HI}$ is the total HI mass
of the host galaxy), specific luminosity ($S_L = 100\times
L_{GCS}/L_V$, where $L_{GCS}$ is the total luminosity of the GCs
and $L_V$ is the luminosity of the host galaxy), specific number
($\hat{T} = 10^9 M_{\odot} \times N_{GC}/M_b$, where $M_b =
M_*+M_{HI}$), globular cluster mass and luminosity normalized
formation efficiencies ($\eta_M, \eta_L$; related to $S_N, S_L,
S_M$ and $\hat{T}$ through equations (23) to (26) of Georgiev et
al. 2010).  According to their model star formation process is
primarily due to stellar and supernovae feed back when the mass is
below $3\times 10^{10} M_{\odot}$ but is governed by virial shock
above this critical mass. \\

We have computed the correlations of some of these parameters with
the tidal index ($\Theta$) for these dwarf galaxies. The
correlations show  very high values for G2 galaxies ($r\sim
0.9/0.8, p< 0.05$, Table 4) contrary to highly insignificant ones
($r \sim 0.09/0.1, p >> 0.05$) for G1 galaxies. This fact
indicates that self enrichment supported by stellar and supernovae
feed back plays a very important role in the formation of stellar
populations in G1 galaxies but star formation is highly regulated
by environment as is evident from high tidal indices, low values
of correlations of $\Theta$ with scaling parameters, low
luminosity - metallicity correlation (Fig. 3) and insignificant
rotation of HI mass for G2 galaxies etc. This might be the result
of globular clusters formation due to higher velocity collisions
in deep potential well leading to more efficient globular cluster
formation. In this respect Kumai et al. (1993) have suggested that
galaxies in deeper environment (i.e. higher $\Theta$) are more
likely to undergo interactions which can increase the random
motion of gas clouds within such galaxies. This leads to increase
in $S_N ( or S_L,S_M, \hat{T_b}, log(\eta_L), log(\eta_m)$ etc)
with environment. At the same time color histograms (Fig. 5) of
GCs as well as of the dwarf galaxies in G2 show major star
formation episode (largest peak) at $<(V-I)_0> \sim 1.0 / 0.9$ for
GC4 and GC1 GCs and $<(V-I)_{e0} \sim 0.9750; $ (viz. Table 4) for
G2 galaxies  which corresponds to older burst of star formation
(viz. Table 3 $\sim Gyr$; Sharina et al. 2008; Puzia \& Sharina
2008). The above phenomenon can be interpreted as star formation
has been ceased subsequently due to gas stripping or ram pressure
sweeping and evaporation which may give rise to different amount
of mass loss as a consequence of the action of the dense
environment. Low HI masses as well as low rotation velocities of
HI masses for G2 galaxies also support the above scenario. This is
in contrast to the low density environment of G1 galaxies which
are free of suffering any external triggering. Hence in low
density environment younger burst of star formation is possible
(Vilchez 1997). This is also clear from the color profiles of GC2
and GC3 GCs (Fig.7) and G1 (Fig. 8) galaxies respectively which
have also peaks at $(V-I) \sim 0.3/0.5$ and those correspond to
age less
than Gyr (Sharina et al. 2005).\\

When globular clusters evolve their core radii decrease and tidal
radii increase. So the quantity $log(r_t/r_c)$ increases. When
$log(r_t/r_c) > $2.5 (Chattopadhyay et al. 2009)the globular
clusters undergo core collapse i.e. they are dynamically much
evolved. Now the values of the above quantity for the four groups
of GCs  GC1, GC2, GC3 and GC4 found as a result of CA are 1.0115,
1.0735, 1.0721 and 1.2483 respectively. So, GCs of GC4 are
dynamically much evolved compared to those in GC1, GC2 and GC3
respectively.  As we know most evolved GCs are roundest so with
respect to ellipticities GCs of GC4 are more evolved than those in
the remaining ones.  So accumulating the above fact and values of
the peaks of the colors in these four groups we can conclude that
GCs of GC4 and GC1  are more evolved than those in GC2 and GC3.\\

The mean values of $(V-I)_0$ for GC1 and GC4 are similar to mean
value of that in G2 whereas the mean values of GC2 and GC3 are
similar to that in G1 (Tables 4 and 5). So G1 galaxies can be
considered as normal sites for the formation of GCs in GC2 and GC3
which are dynamically less evolved (viz. $<Age> \sim$ 5 Gyr for
GC4, Table 5). On the other hand G2 galaxies can be considered as
places for the formation of of GCs in GC1 and GC4 which are
dynamically much evolved hence older (age $\sim$ 7.2 Gyr for GC3,
Table 5). Though ellipticities and $log(r_t/r_c)$ for GC1 do not
support the above fact but the higher mean value of color
$(V-I)_0$ indicates that it contains redder GCs which is an
indication of older ages. The GCs in GC1 and GC4 are formed by
mechanism other than self enrichment (viz. $r(V0, (V-I)_0) \sim
-0.199, p = 0.350$ for GC1 and $r(V0, (V-I)_0) \sim -0.155, p =
0.49$ for GC4). From the histograms of colors (Fig. 7) and color
vs projected distance (Fig. 9) of the groups GC1, GC2, GC3 and GC4
it is clear that the highest peaks of GCs of GC1 and GC4 occur at
higher values of $(V-I)_0$. But for GCs in GC2 and GC3 the heights
of the peaks at the modes are not very different from one another
and they occur even at much lower values of $(V-I)_0$ (viz.
0.3/0.5, Table 5). If the correlations between color and projected
distance are calculated for the four groups GC1, GC2, GC3 and GC4
these are (0.102, p = 0.635), (0.402, p = 0.109), (0.234, p =
0.164) and (0.091, p= 687) respectively. So it is clear that for
GCs of GC2 and GC3 the correlations are
moderate at 10 \% level of significance . \\

The correlations between magnitude and projected distance of the
above four groups are (0.015, p = 0.945), (0.560, p = 0.019),
(0.373, p = 0.023) and ( -0.180, p = 0.422) respectively.  This is
an indication that the star formation history can be considered to
be similar for groups GC1 and GC4 compared to that for GC2 and
GC3.  Since the GCs in GC1 and GC4 are much evolved than those in
GC2 and GC3, and the tidal indices are higher for those in G2
which are their places for formation, it is very likely to assume
that the GCs in the outer parts of GC1 and GC4 are tidally
stripped from their host galaxies. This does not hold for GCs in
GC2 if their host galaxies (considered G1) possess high degree of
rotation which is the case as discussed before regarding the
rotation of their total HI
masses though this is not true for GCs of GC3\\

\section{Summary and conclusions}

In the present work statistical approach for classification of LSB
dwarf galaxies and globular clusters has been developed. For the
classification two statistical techniques are used viz. Principal
Component Analysis (PCA) followed by K-means Cluster Analysis (CA)
together with the criterion for finding optimum number of
homogeneous groups. Through PCA the optimum set of parameters
giving maximum variation among the objects is found while the
required homogeneous groups are found using CA. The optimum number
of groups is found following Sugar \& James (2003). For the sample
of dwarf galaxies two groups are found primarily indicative of
their masses ($M_{V0}$), tidal indices ($\Theta$)and surface
brightness averaged over effective radius ($SBV_{e0}$) but
irrespective of their morphological indices. \\

G1 galaxies are massive ($M_{V0} < -13$) with larger amount of HI
mass having higher degree of rotation ($V_m$) and lower mean value
of tidal index with absence of any correlation ($r \sim 0.09/0.1,
p>>0.05$; viz Table 4) with scaling parameters. Also there exists
moderate mass metallicity correlations among the dwarf galaxies of
G1. All these facts indicate that dwarf galaxies of G1 are formed
by self enrichment supported by stellar and supernovae feedback.
On the other hand G2 galaxies are less massive, have insignificant
amount of HI mass with little or absence of any rotation , devoid
of any mass metallicity correlation, high values of tidal indices
with significant correlations ($r \sim 0.9/0.8, p < 0.05$; viz.
Table 4) with the scaling parameters. The above mentioned
characteristics suggest that environment plays a very important
role in formation in the star formation scenario of
these dwarf galaxies.\\

Subsequently a classification of GCs in the LV has been carried
out and four groups emerged as a result of such classification. A
comparison of the color profiles of the GCs in these groups with
those of dwarf galaxies suggest that among the four groups GC1 and
GC4 which are dynamically much evolved  can be formed in G2
whereas dynamically less evolved GCs in GC2 and GC3 having no
significant self-enrichment can be formed in galaxies like G1.
Also colors of GCs in GC1 and GC4 bear no correlations with their
projected distances while moderate correlation exists for the GCs
of GC2 and GC3. This is also true for magnitude vs projected
distance correlations. So the star formation history for the GCs
of GC1 and GC4 might be speculated to be different from those in
GC2 and GC3.

\clearpage
\section{Acknowledgements}

One of the the authors (Tanuka Chattopadhyay)wishes to thank
Department of Science and Technology (DST),India for awarding her
a major research project for the work. The authors are grateful to
Prof A.K. Chattopadhyay and Emmanuel Davoust for their useful
suggestions and help. The authors are also thankful for the
suggestions of the referee.

\clearpage

\begin{figure}
\epsscale{.80}\plotone{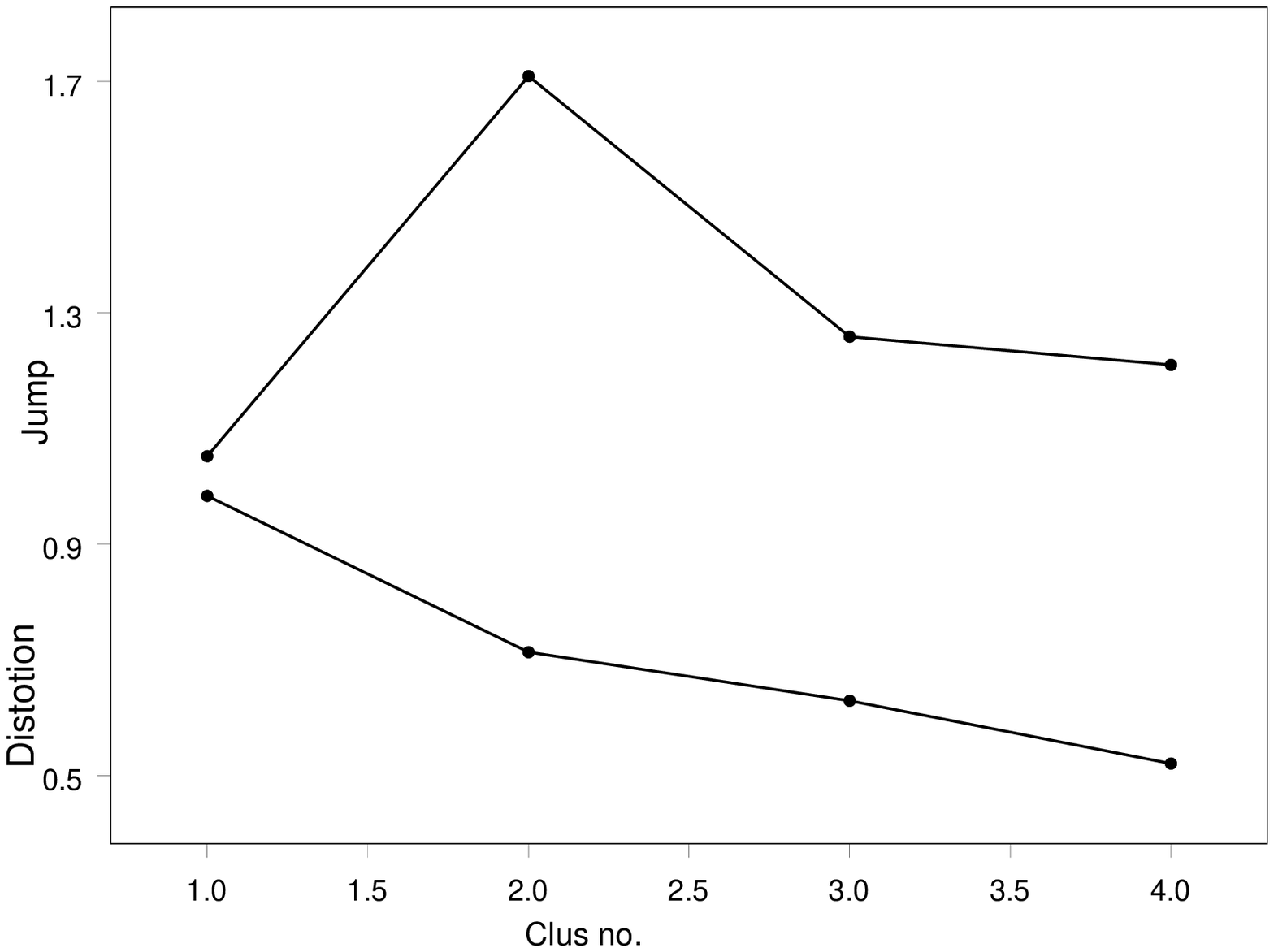} \caption{The distortion and jump
curves for the classification of dwarf galaxies in the LV}
\end{figure}

\clearpage

\begin{figure}
\epsscale{.80} \plotone{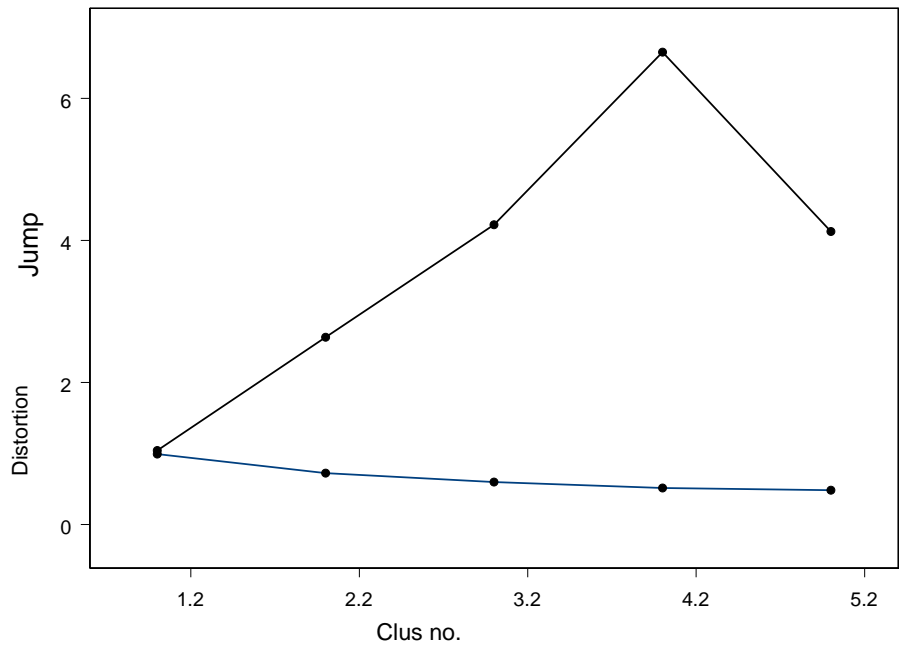} \caption{The distortion and jump
curves for the classification of globular clusters in the LV}
\end{figure}

\clearpage

\begin{figure}
\begin{center}
\includegraphics[angle=-90,width=15cm]{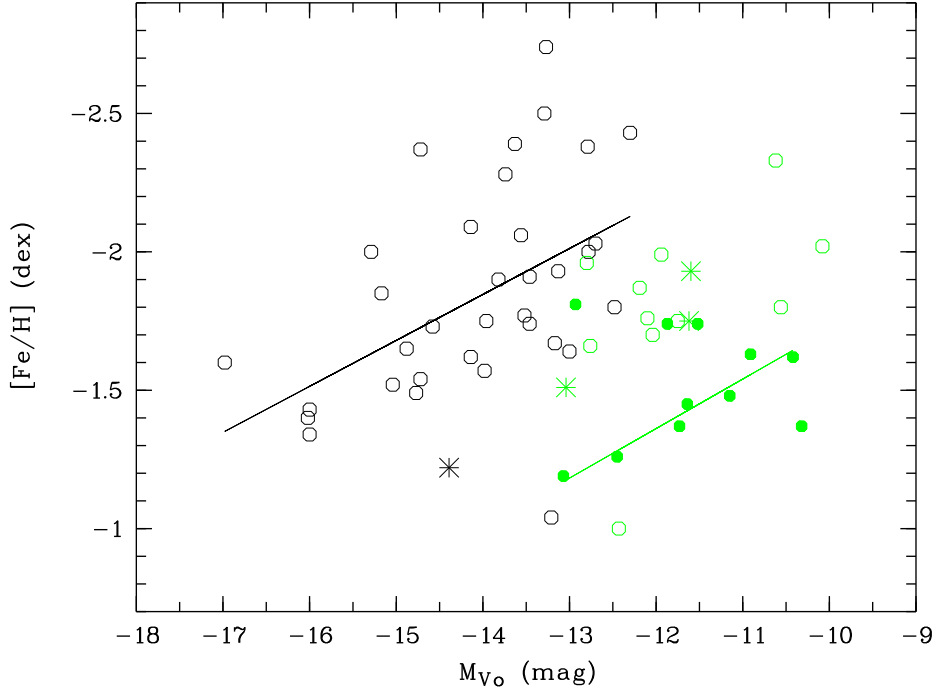}
\caption{Absolute luminosity in the V-band($M_{V0}$) vs.
 metallicity ($[Fe/H]$) diagram for the two groups G1 and G2 of dwarf galaxies found
as a result of CA in the LV. The black circles (dIrrs) and one
black diamond (dIrr/dSph) are for group G1 and green symbols are
for G2. dSphs, dIrrs, and dIrrs/dSphs are shown as dots, circles,
and asterisks, correspondingly. The best fitted line is for G1
galaxies removing top 2 black circles as outliers. The green line
is from S08.}
\end{center}
\end{figure}

\clearpage

\clearpage

\begin{figure}
\epsscale{.80}
\begin{center}
\includegraphics[angle=-90,width=15cm]{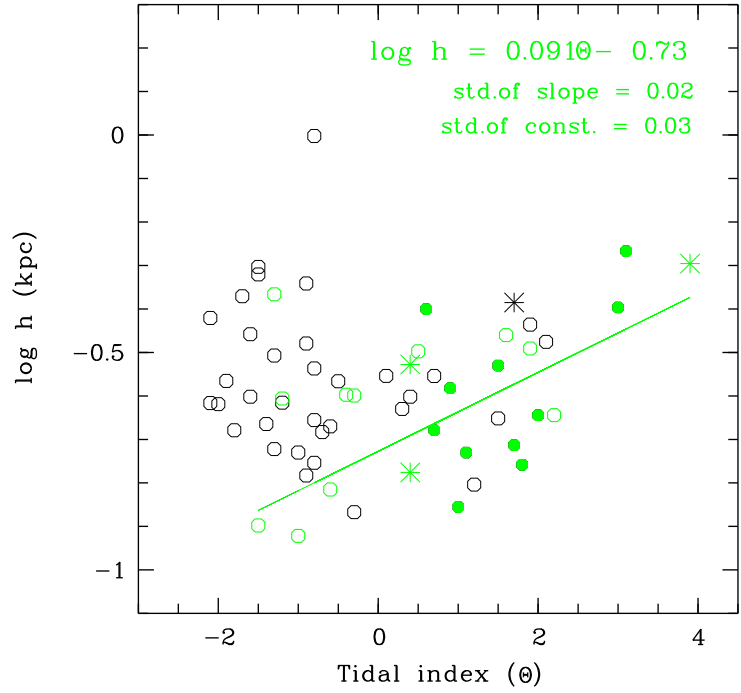}
\caption{Tidal index ($\Theta$) vs logarithm of scale length in
kpc for the two groups G1 and G2 of dwarf galaxies found as a
result of CA in the LV. Symbols are the same as in Fig.3.
Regression line for G2 was counted removing E443-09 and KKH5 as
outliers.}
\end{center}
\end{figure}
\clearpage

\begin{figure}
\epsscale{.80}
\begin{center}
\includegraphics[angle=-90,width=15cm]{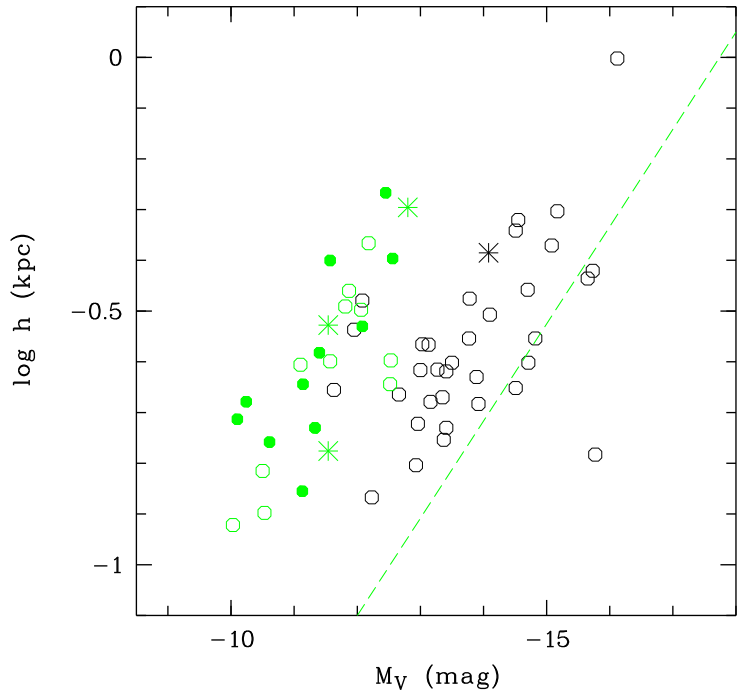}
\caption{Absolute magnitude in the V-band, corrected for Galactic
extinction vs. projected distance of a GC from a center of a
galaxy for the two groups G1 and G2 of dwarf galaxies found as a
result of CA in the LV. Symbols are the same as in Fig.3. Dotted
line is a line of equal central surface brightness for spiral
galaxies from S08.}
\end{center}
\end{figure}
\clearpage

\begin{figure}
\epsscale{.80} \plotone{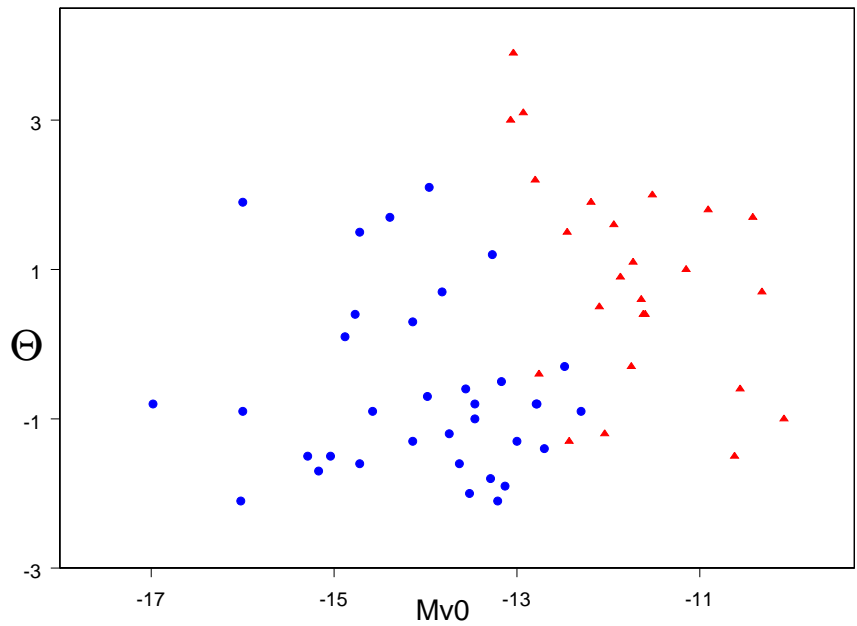} \caption{Tidal index ($\Theta$) vs
Mass($M_{V0}$) diagram for the two groups G1 and G2 of dwarf
galaxies found as a result of CA in the LV. The blue solid circles
are for group G1 and red solid triangles are for G2}
\end{figure}

\clearpage

\begin{figure}
\epsscale{.80} \plotone{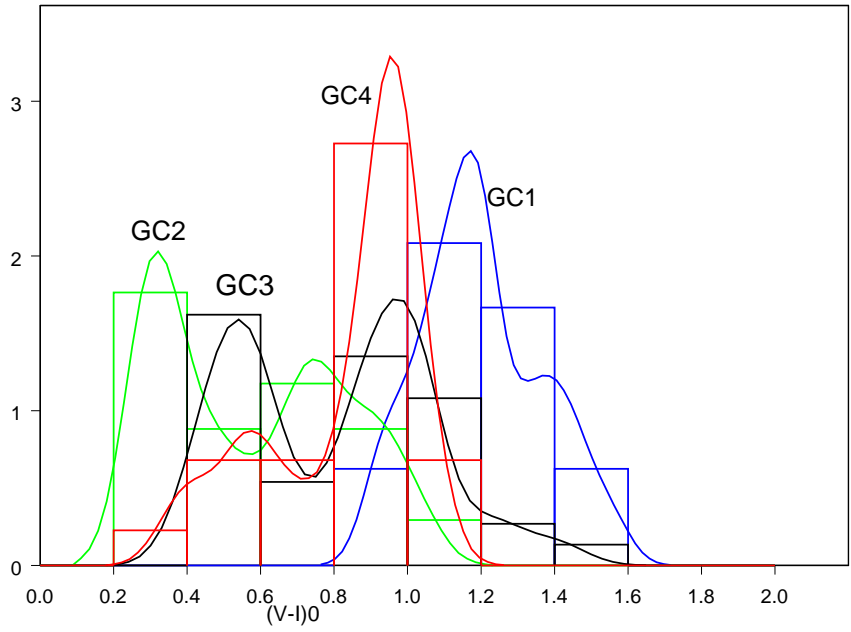} \caption{Histograms with density
lines of $(V-I)_0$ color for the groups of globular clusters found
as a result of CA in the LV. Blue solid curve is for GC1, green
solid curve is for GC2, black solid curve is for GC3 and red solid
curve is for GC4 respectively. }
\end{figure}

\clearpage

\begin{figure}
\epsscale{.80} \plotone{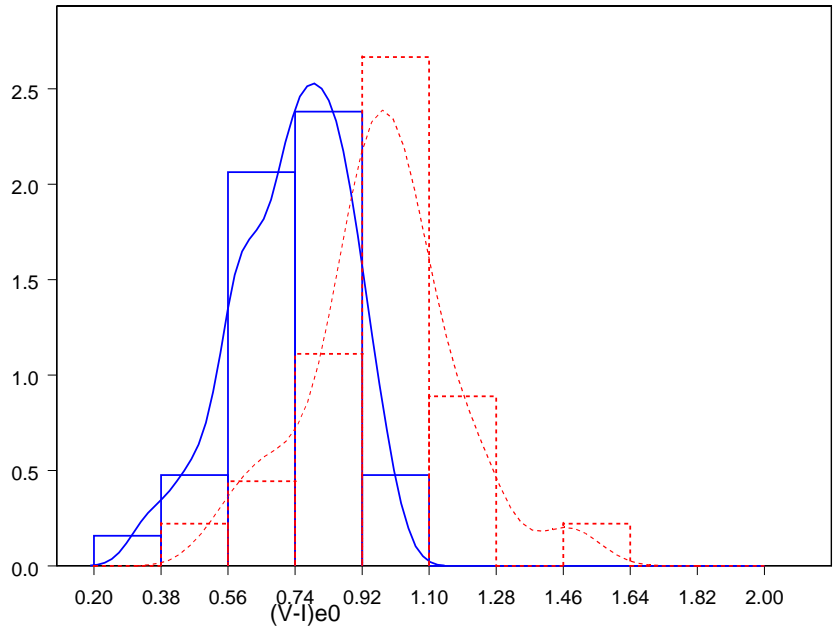} \caption{Histograms with density
lines of $(V-I)_{e0}$ color for the groups of dwarf galaxies found
as a result of CA in the LV. Blue solid line is for G1 galaxies
and red dashed line is that for G2.}
\end{figure}

\clearpage

\begin{figure}
\epsscale{.80} \plotone{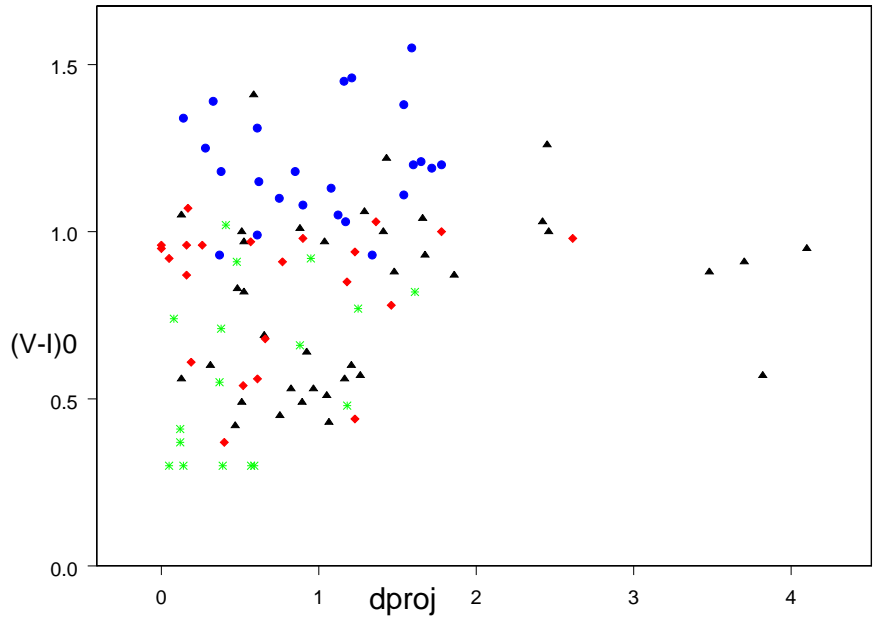} \caption{Color ($(V-I)_0$) vs
projected distances ($d_{proj}$) of the four groups of GCs GC1,
GC2, GC3 and GC4. Blue solid circles are for GC1, green stars  are
for GC2, black triangles are for GC3 and red diamonds  are for
GC4.}
\end{figure}

\clearpage

\begin{figure}
\epsscale{.80} \plotone{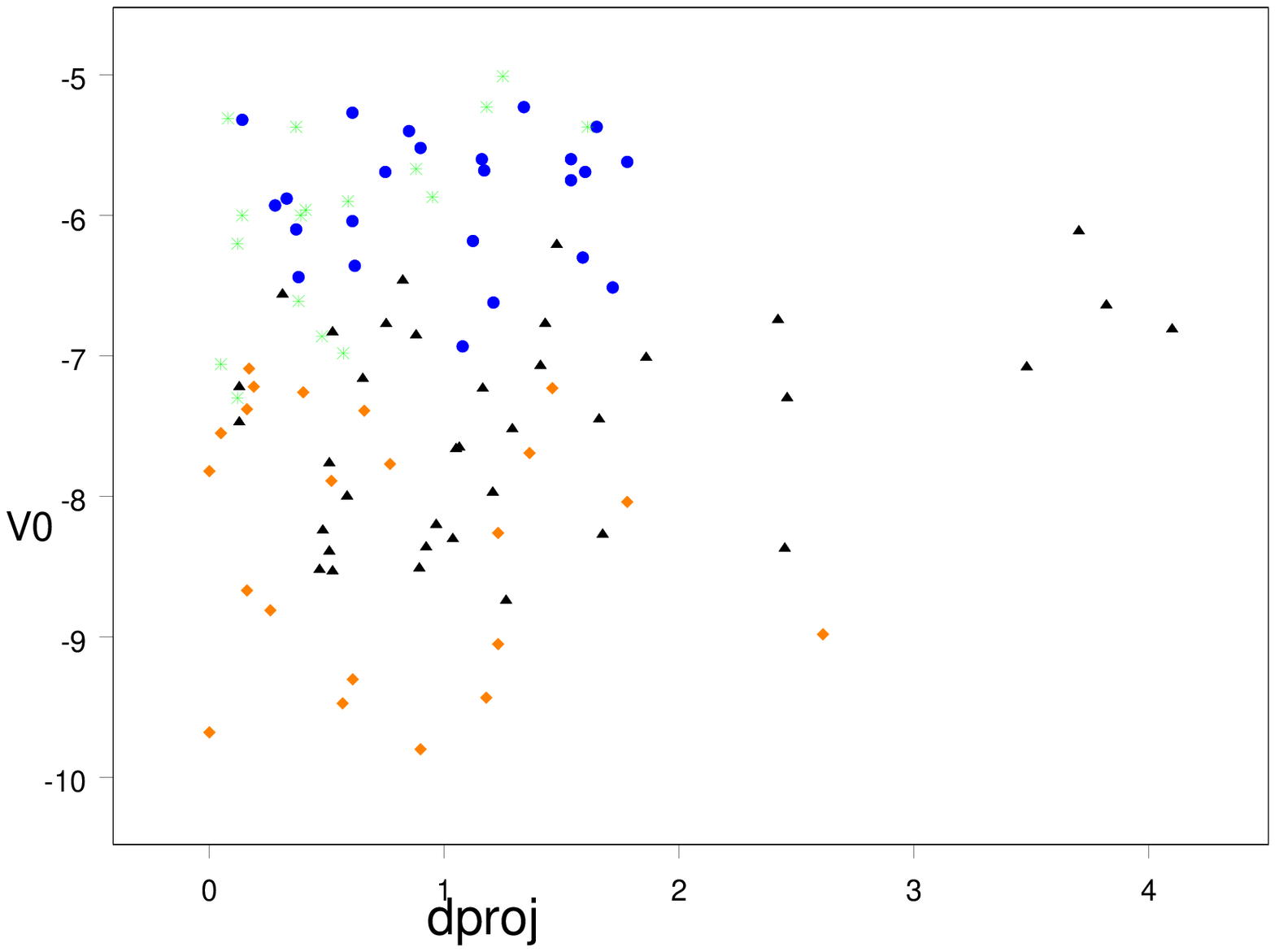} \caption{Absolute magnitude V0 vs
$d_{proj}$ for the groups of GCs GC1, GC2, GC3 and GC4. The
symbols and colors are same as Fig. 9.}
\end{figure}

\clearpage

\begin{deluxetable}{ccc}
\tabletypesize{\scriptsize} \tablecaption{List of  dwarfs in Data
Set 1} \tablewidth{0pt} \tablehead{ \colhead{Name}&
\colhead{RA(2000)}& \colhead{DEC(2000)}} \startdata
 E349-031 &00 \ 08 \ 13.3 &-34 \ 34 \ 42.0 \\
 E410-005 &00 \ 15 \ 31.4 &-32 \ 10 \ 48.0 \\
E294-01 &00 \ 26 \ 33.3 &-41 \ 51 \ 20.0\\
KDG2&00 \ 49 \ 21.1 &-18 \ 04 \ 28.0 \\
E540-032&00 \ 50 \ 24.3 &-19 \ 54 \ 24.0 \\
UGC685&01 \ 07 \ 22.3 &16 \ 41 \ 02.0 \\
KKH5&01 \ 07 \ 32.5 &51 \ 26 \ 25.0 \\
KKH6&01 \ 34 \ 51.6 &52 \ 05 \ 30.0 \\
KK16&01 \ 55 \ 20.6 &27 \ 57 \ 15.0 \\
KK17&02 \ 00 \ 09.9 &28 \ 49 \ 57.0 \\
KKH34&05 \ 59 \ 41.2 &73 \ 25 \ 39.0 \\
E121-20&06 \ 15 \ 54.5 &-57 \ 43 \ 35.0 \\
E489-56&06 \ 26 \ 17.0 &-26 \ 15 \ 56.0 \\
KKH37&06 \ 47 \ 45.8 &80 \ 07 \ 26.0 \\
UGC3755&07 \ 13 \ 51.8 &10 \ 31 \ 19.0 \\
E059-01&07 \ 31 \ 19.3 & -68 \ 11 \ 10.0\\
KK65&07 \ 42 \ 31.2 &16 \ 33 \ 40.0 \\
UGC4115&07 \ 57 \ 01.8 &14 \ 23 \ 27.0 \\
DDO52&08 \ 28 \ 28.5 &41 \ 51 \ 24.0 \\
D564-08&09 \ 02 \ 54.0 &20 \ 04 \ 31.0  \\
D565-06&09 \ 19 \ 29.4 &21 \ 36 \ 12.0 \\
KDG61&09 \ 57 \ 02.7 &68 \ 35 \ 30.0 \\
KKH57&10 \ 00 \ 16.0 &63 \ 11 \ 06.0 \\
HS117&10 \ 21 \ 25.2 &71 \ 06 \ 58.0 \\
UGC6541&11 \ 33 \ 29.1 &49 \ 14 \ 17.0 \\
NGC3741&11 \ 36 \ 06.4 &45 \ 17 \ 07.0 \\
E320-14&11 \ 37 \ 53.4 &-39 \ 13 \ 14.0 \\
KK109&11 \ 47 \ 11.2 &43 \ 40 \ 19.0 \\
E379-07&11 \ 54 \ 43.0 &-33 \ 33 \ 29.0 \\
NGC4163&12 \ 12 \ 08.9 &36 \ 10 \ 10.0  \\
UGC7242 &12 \ 14 \ 07.4 &66 \ 05 \ 32.0 \\
DDO113&12 \ 14 \ 57.9 &36 \ 13 \ 08.0 \\
DDO125&12 \ 27 \ 41.8 & 43 \ 29 \ 38.0\\
UGC7605&12 \ 28 \ 39.0 &35 \ 43 \ 05.0 \\
E381-018&12 \ 44 \ 42.7 & -35 \ 58 \ 00.0 \\
E443-09 &12 \ 54 \ 53.6 &-28 \ 20 \ 27.0 \\
KK182&13 \ 05 \ 02.9 &-40 \ 04 \ 58.0 \\
UGC8215&13 \ 08 \ 03.6 &46 \ 49 \ 41.0 \\
E269-58 & 13 \ 10 \ 32.9& -46 \ 59 \ 27.0\\
KK189&13 \ 12 \ 45.0 &-41 \ 49 \ 55.0  \\
E269-66&13 \ 13 \ 09.2 &-44 \ 53 \ 24.0 \\
KK196&13 \ 21 \ 47.1 &-45 \ 03 \ 48.0 \\
KK197&13 \ 22 \ 01.8 &-42 \ 32 \ 08.0 \\
KKs55& 13 \ 22 \ 12.4& -42 \ 43 \ 51.0\\
14247&13 \ 26 \ 44.4 &-30 \ 21 \ 45.0 \\
UGC8508&13 \ 30 \ 44.4 &54 \ 54 \ 36.0 \\
E444-78&13 \ 36 \ 30.8 &-29 \ 14 \ 11.0 \\
UGC8638&13 \ 39 \ 19.4 &24 \ 46 \ 33.0 \\
KKs57&13 \ 41 \ 38.1 &-42 \ 34 \ 55.0 \\
KK211&13 \ 42 \ 05.6 &-45 \ 12 \ 18.0  \\
KK213&13 \ 43 \ 35.8 &-43 \ 46 \ 09.0 \\
KK217&13 \ 46 \ 17.2 &-45 \ 41 \ 05.0 \\
CenN&13 \ 48 \ 09.2 &-47 \ 33 \ 54.0 \\
KKH86&13 \ 54 \ 33.6 &04 \ 14 \ 35.0 \\
UGC8833&13 \ 54 \ 48.7 &35 \ 50 \ 15.0  \\
E384-016&13 \ 57 \ 01.6 &-35 \ 20 \ 02.0 \\
KK230&14 \ 07 \ 10.7 & 35 \ 03 \ 37.0\\
DDO190 &14 \ 24 \ 43.5 &44 \ 31 \ 33.0 \\
E223-09&15 \ 01 \ 08.5 &-48 \ 17 \ 33.0 \\
IC4662&17 \ 47 \ 06.3 &-64 \ 38 \ 25.0  \\ \hline
\enddata
\end{deluxetable}
\clearpage

\begin{deluxetable}{ccc}
\tabletypesize{\scriptsize} \tablecaption{Discriminant analysis
for dwarf galaxies in LV. G1, G2 are the groups found by K-means
and $G1^*$ and $G2^*$ are the groups to which dwarfs are assigned
by the Discrminant Analysis. N = 60, $N_{correct}$ = 59,
Proportion correct = 0.983} \tablewidth{0pt} \tablehead{
\colhead{} & \colhead{}& \colhead{Number of members}} \startdata
 DA Clusters &G1&G2 \\\hline
     $G1^*$&35 &1 \\
     $G2^*$&0 & 24\\ \hline
     Total &35& 25 \\ \hline

\enddata
\end{deluxetable}

\begin{deluxetable}{ccccc}
\tabletypesize{\scriptsize} \tablecaption{Discriminant analysis
for GCs in the the LV dwarf galaxies. GC1, GC2, GC3 and GC4  are
the groups found by K-means and $GC1^*$, $GC2^*$, $GC3^*$ and
$GC4^*$ are the groups to which GCs are assigned by the
Discrminant Analysis. N = 100, $N_{correct}$ = 97, Proportion
correct = 0.97} \tablewidth{0pt} \tablehead{ \colhead{} &
\colhead{}&\colhead{}& \colhead{No.}&\colhead{}}\startdata
 DA Clusters &GC1&GC2 & GC3& GC4\\\hline
     $GC1^*$&23 &0 & 0 & 0\\
     $GC2^*$&0 & 17& 0& 1 \\
     $GC3^*$&1 & 0&37& 1 \\
     $GC4^*$& 0& 0& 0& 20 \\ \hline
     Total &24& 17& 37 & 22 \\ \hline

\enddata
\end{deluxetable}

\clearpage

\begin{deluxetable}{ccc}
\tabletypesize{\scriptsize} \tablecaption{Mean values and
correlations of the significant parameters for the two groups of
dwarfs in the LV} \tablewidth{0pt} \tablehead{
\colhead{Parameters} & \colhead{G1}& \colhead{G2}} \startdata
 Number&35&25 \\
     $\Theta$&-0.631 $\pm$0.200 &0.880 $\pm$0.287 \\
     $[Fe/H]$&-1.8394 $\pm$0.0658 & -1.6675 $\pm$ 0.0586 \\
     $M_{V0}$&-14.060 $\pm $0.190 &-11.742 $\pm$ 0.177 \\
     $SBV_{e0}$ & 22.730 $\pm$ 0.150 & 24.102 $\pm$ 0.128\\
     $(V-I)_{e0}$ & 0.7298 $\pm$ 0.0257 & 0.9750 $\pm$ 0.0391 \\
     $logR_e$ & -0.2827 $\pm$ 0.0331& -0.3842 $\pm$ 0.0357\\
     $V_m$ & 26.12 $\pm$ 3.23 & 11.11 $\pm$ 3.69 \\
     $M_{HI}$ & 6.571 $\pm$ 2.40 & 0.516 $\pm$ 0.249\\
     $M_{*V}$ & 6.69 $\pm$ 3.06 & 0.41$\pm$ 0.155 \\
     $\hat{T_b}$ & 52.6 $\pm$ 15.7 & 238.1 $\pm$ 96.4 \\
     $log(\eta_L)$ & -4.264 $\pm$ 0.224 & -4.568 $\pm$ 0.329 \\
     $ log(\eta_M)$ & -4.616 $\pm$ 0.301 & -4.912 $\pm$ 0.47 \\
     Correlations &r \ \ \ p & r \ \ \ p\\
     $(log\eta_L, \Theta)$ & 0.096\ \ \ 0.805  & 0.987 \ \ \ 0.002\\
     $(log\eta_M, \Theta)$ & 0.345 \ \ \ 0.364 & 0.956 \ \ \ 0.011\\
     $(S_L,\Theta)$ & 0.102 \ \ \ 0.793& 0.889 \ \ \ 0.044\\
     $(S_M,\Theta)$ &0.200 \ \ \ 0.606& 0.896\ \ \ 0.040\\
     $(\hat{T_b}, \Theta)$ & 0.460 \ \ \ 0.213& 0.854 \ \ \ 0.065\\
     $([Fe/H], M_v)$ & -0.553 \ \ \ 0.001 &-0.290 \ \ \ 0.160  \\
     $(M_{V0}, (V-I)_{e0}) $& -0.224 \ \ \ 0.195 & -0.155 \ \ \ 0.459\\
\enddata
\end{deluxetable}
\clearpage

\begin{deluxetable}{ccccc}
\tabletypesize{\scriptsize} \tablecaption{Mean values of the
significant parameters for the three groups of GCs in the LV}
\tablewidth{0pt} \tablehead{ \colhead{Parameters} & \colhead{GC1}&
\colhead{GC2}& \colhead{GC3}& \colhead{GC4}} \startdata

 Number &24 & 17& 37& 22\\
 $ V0$ & -5.8765 $\pm$ 0.0954& -6.041 $\pm$ 0.170& -7.481 $\pm$0.123& -8.263 $\pm$ 0.194 \\
 $\mu_{V0}$ & 21.075 $\pm$ 0.097& 19.903 $\pm$ 0.140& 20.443 $\pm$ 0.141& 18.127 $\pm$ 159\\
 $log(r_h)$ & 0.8863 $\pm$ 0.0235& 0.7024 $\pm$ 0.0287& 1.0753 $\pm$ 0.0153& 0.7820 $\pm$ 0.0321\\
 $log(r_t)$ & 1.5556 $\pm$ 0.0578& 1.4077 $\pm$ 0.0672& 1.7965 $\pm$0.0557& 1.6179 $\pm$ 0.0463 \\
 $log(r_c)$ & 0.5441 $\pm$ 0.0366& 0.3342 $\pm$ 0.0403& 0.7244 $\pm$ 0.0197& 0.3696 $\pm$ 0.0377\\
 $log(r_t/r_c)$ & 1.0115 $\pm$ 0.0747& 1.0735 $\pm$ 0.0873& 1.0721 $\pm$ 0.0625& 1.2483 $\pm$ 0.0500\\
e           & 0.1542 $\pm$ 0.0208& 0.1647 $\pm$ 0.0284& 0.1081 $\pm $0.0166 & 0.10$\pm$ 0.01\\
 $(V-I)_0$ & 1.1996 $\pm$ 0.0340& 0.5800 $\pm$ 0.0617& 0.8035 $\pm$ 0.0430& 0.8332$\pm$ 0.0436\\
 $d_{proj}$ & 1.014 $\pm$ 0.105& 0.563 $\pm$ 0.112& 1.354 $\pm$ 0.171& 0.740 $\pm$ 0.144\\
 $Age(Gyr)$ & *& *               & 5.0$\pm$1.32 & 7.2 $\pm$  1.5 \\
 $Z/H$ & *&*  & -1.167 $\pm$ 0.230& -1.3 $\pm$0.307 \\
 $(\alpha/ H) $ & *&* & 0.2 $\pm$0.0632& 0.18 $\pm$0.0490 \\

\enddata
\end{deluxetable}
\end{document}